\documentclass[prd,aps,twocolumn]{revtex4}
\usepackage{amsmath}
\usepackage{graphicx}
\usepackage{amssymb}
\usepackage{hyperref}
\usepackage{color}

\definecolor{verde}{rgb}{0,0.5,0}

\def\be{\begin{equation}}
\def\ee{\end{equation}}
\def\bea{\begin{eqnarray}}
\def\eea{\end{eqnarray}}
\def\be{\begin{equation}}
\def\ee{\end{equation}}
\def\ba{\begin{align}}
\def\ea{\end{align}}
\def\p{\partial}
\def\noi{\noindent}

\newcommand\kv{k_{\rm V}}
\newcommand\knl{k_{\rm NL}}

\newcommand\lsim{\mathrel{\rlap{\lower4pt\hbox{\hskip0.5pt$\sim$}}
    \raise1pt\hbox{$<$}}}
\newcommand\gsim{\mathrel{\rlap{\lower4pt\hbox{\hskip0.5pt$\sim$}}
    \raise1pt\hbox{$>$}}}

\begin{document}
%\preprint{}
%\draft

%
% Remove this and closure after abstract, plus preprint number,
% in electronic submission
%
\input epsf
\renewcommand{\topfraction}{0.99}
\renewcommand{\bottomfraction}{0.99}
%\twocolumn[\hsize\textwidth\columnwidth\hsize\csname@twocolumnfalse\endcsname]

\title{Screening in perturbative approaches to LSS}
\author{Matteo Fasiello$^a$, Zvonimir Vlah$^{a,b}$}

\affiliation{$^a$Stanford Institute for Theoretical Physics and Department of Physics, Stanford University, Stanford, CA 94306}
\affiliation{$^b$Kavli Institute for Particle Astrophysics and Cosmology, Stanford University and SLAC, Menlo Park, CA 94025 }

\begin{abstract}
\noi A specific value for the cosmological constant $\Lambda$ can account for late-time cosmic acceleration. However, motivated by the so-called cosmological constant problem(s), several alternative mechanisms have been explored. To date, a host of well-studied dynamical dark energy and modified gravity models exists. Going beyond $\Lambda$CDM often comes with additional degrees of freedom (dofs). For these to pass existing observational tests, an efficient screening mechanism must be in place. The linear and quasi-linear regimes of structure formation are ideal probes of such dofs and can capture the onset of screening. We propose here a semi-phenomenological ``filter" to account for screening dynamics on LSS observables, with special emphasis on Vainshtein-type screening. 
\end{abstract}

%%%%%%%%%%%%%%%%%%%%%%%%%%%%%%%%%%%%%%%%%%%%%%%%%%%%%%%%%%%%%%%%%%%%%%%%
\maketitle
%\section{Introduction}
\noindent
\noindent

\section {Introduction}
\label{intro}

 The existence of a dynamical mechanism responsible for late-time cosmic acceleration often requires additional degrees of freedom (dofs) besides those of general relativity. On the other hand, the latter is, to an exquisite level of accuracy, a good description of the physics we see at ``small" scales such as within the solar system. For the overall picture to be consistent, a screening mechanism must be in place. Screening is expected to be efficient in highly dense regions. Conversely, low-density environments make up the ideal settings to access the additional dynamics of beyond-$\Lambda$CDM models. 

Large scale structure probes are an optimal case in point.  The linear regime of structure formation is the environment where the additional dofs are most transparent and testable. These scales are well-described by perturbation theory. Crucially, the number of available modes grows approximately like the cube of the wavenumber, making any gain on the k-reach of the perturbative theory significant. An analytical description of the mildly-non-linear regime of structure formation \cite{Bernardeau:2001qr,Baumann:2010tm,Porto:2013qua} is highly desirable: these scales are a precious repository of information on both primordial physics (e.g. non-Gaussianities \cite{Dalal:2007cu,Angulo:2015eqa,Assassi:2015fma}) and late-time dynamics (see \cite{Huterer:2013xky} and references therein).
Our focus here will be on the latter: the mildly-non-linear regime can capture the onset of screening dynamics, which is central to dark energy and modified gravity models. 

\section{A new scale}
\label{scale}
 There has been considerable recent effort towards expanding the fluid description of dark matter to include an additional dynamical component  (see e.g. \cite{Fasiello:2016qpn,Fasiello:2016yvr} and  \cite{Sefusatti:2011cm,Anselmi:2011ef,Anselmi:2014nya} for earlier work on the same specific model). These works are based on the notion that the large hierarchy of scales in between the size of the observable universe $1/H_0$ and the highly non-linear-regime of structure formation $1/k_{\rm NL}$ allows for a clean perturbative treatment of the $k\ll k_{NL}$ modes. Naturally, the small expansion parameter is  $k/k_{\rm NL}$. 
By employing a full-fledged effective theory approach \cite{Baumann:2010tm,Carrasco:2012cv}, the microphysics of yet smaller scales can be encapsulated in a number of ``UV" coefficients \footnote{These multiply at each order all possible operators allowed  by the symmetries of the theory (e.g. rotational invariance).} to be determined by comparison with observations and/or simulations.

 However, as argued above, in general screening will suppress the effects of the additional dofs in dark-energy (DE) and modified gravity (MG) at small scales i.e. in the highly-non-linear regime. We sketch in Fig.~(1) the \textit{total} power spectrum $vs$ the $\Lambda$CDM behaviour under one of the screening mechanisms that most clearly exemplifies this effect: Vainshtein screening  (see e.g. \cite{Falck:2014jwa} for a detailed N-body analysis).

\begin{figure*}[t!]
\label{fig1}
\includegraphics[scale=0.7]{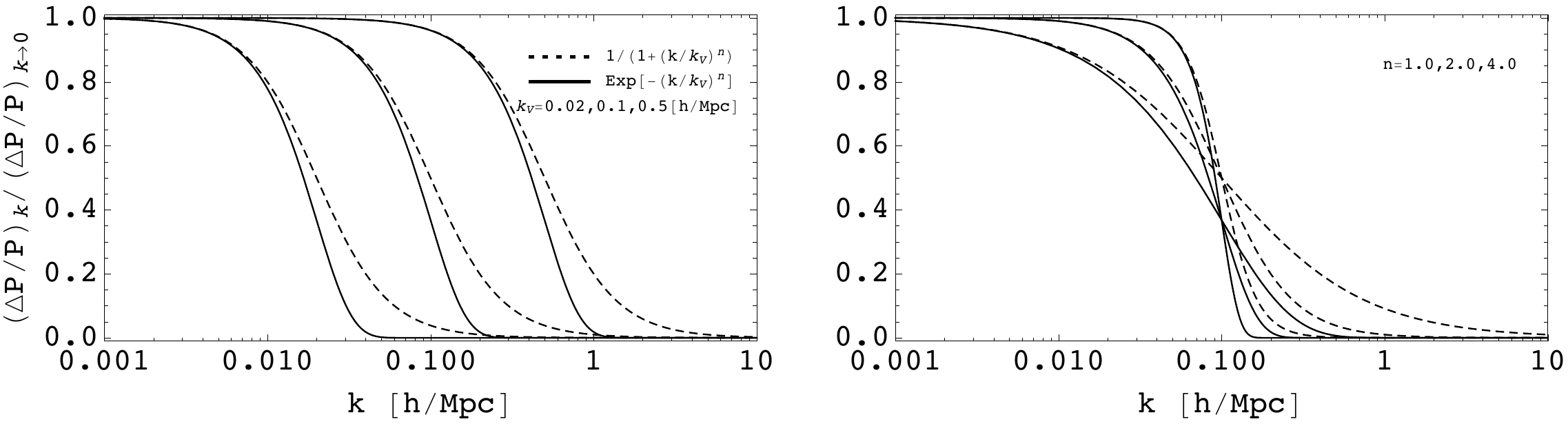}
\caption{Shown in this plot is a sketch of screening effects on the fractional difference between the power spectrum in nDGP \cite{Dvali:2000hr} (here as a Vainshtein-screened theory and a precursor of Galileons and massive gravity theories) and $\Lambda$CDM. 
Two test functional forms are used: Lorentzian (dashed line) and Gaussian-like (solid line), for several different values of the parameter  $k_{\rm_V}$ and slope $n$. We refer the reader to the discussion around Eq.~(\ref{kernels}) for more details.
 It's clear that, according to the region of parameter space probed (different  $k_{\rm V}$ values), screening can happen already at linear scales, at quasi-linear ones, or only deep in the non-linear regime. We refer the reader to Fig.~(1) of \cite{Falck:2014jwa} for the results, derived via N-body simulations. Note also that here, as opposed to \cite{Falck:2014jwa}, we have normalized the profiles by their value at low $k$.}
\end{figure*}

 There exists in other words a scale, we shall call it $k_{\rm V}$, at which screening becomes active. Any attempt at an accurate and general description of beyond-$\Lambda$CDM dynamics of structure formation in screened theories needs to take $k_{\rm V}$ into account (see \cite{Alonso:2016suf} for interesting work that includes $k_{\rm V}$-related effects up to linear order). In the specific case of Vainshtein-screened theories, some readers may be more familiar with the quantity in coordinate space related to $k_{\rm V}$, the so-called \textit{Vainshtein} radius, $r_{\rm V}$. Models such as DGP \cite{Dvali:2000hr} and non-linear massive gravity \cite{deRham:2010kj} exhibit an efficient implementation of such screening mechanism.

 The definition of $r_{\rm V}$ typically depends on the specific configuration \footnote{For example, the effective dimensionality and symmetry of the source + test-particle system.} under study. Most importantly, it depends on a set of defining parameters for the theory. For the above examples, $r_{\rm V}$ depends on the \textit{cross-over scale} in DGP and on the \textit{graviton mass} in massive gravity. It is then clear how the mildly-non-linear regime of stucture formation can be used to set powerful bounds on DE and MG models.

 Our ability to access screening depends crucially on the $k_{\rm NL}$ \textit{vs} $k_{\rm V}$ hierarchy:\\ 
\noi - In the asymptotic region where $k_{\rm V}\gg k_{\rm NL}$ accounting for screening is hardly necessary: all dofs are manifest and the perturbative expansion breaks down (at $k \lesssim k_{\rm NL}$) long before screening becomes relevant.\\
\noi - Complementarily, for too small a $k_{\rm V}$, $k_{\rm V}\ll k_{\rm NL}$,  screening will be extremely efficient and for all intents and purposes our description will coincide with $\Lambda$CDM.\\ 
 The interesting regime at hand corresponds to 
 \bea k_{\rm V} \lesssim k_{\rm NL} \; .
 \eea
 Intriguingly, there exist several setups where this regime provides the most compelling cosmological solutions (see \cite{DAmico:2011eto} for one such example).

\subsection*{Useful mismatch in the two expansions}
 As shown in Fig.(1), depending on the $\kv \,{\rm vs}\,\, \knl$ hierarchy, screening can become relevant already at linear scales, or only at $n$-loop order in the $k/k_{\rm NL}$ expansion, or in general ``in between" loops. This happens because non-linearities can become important at very different scales on the dark matter and the dark energy side (see also Fig.~(1) of \cite{Falck:2014jwa}). It is relying on this very fact that one can hope to access an intrinsically non-linear (in $k/k_{\rm V}$ on the MG side) phenomenon such as screening already at quasi linear (in $k/k_{\rm NL}$ on the DM side) scales. 

Depending on the value of $k_{\rm V}$ and the strength of screening, non-linearities on the MG side can start suppressing the gravitational coupling between MG and DM at very different scales. It is this allowed ``mismatch" between $k_{\rm V}$ and $k_{\rm NL}$ that grants access to screening. Our screening model will be a phenomenological take on highly non-linear screening effects for the power spectrum (PS) of the total density contrast: it should be thought of as resulting from the resummation of the non-linearities in $k/k_{\rm V}$ all the while the perturbative expansion is kept for the $k/k_{\rm NL}$ parameter.

\section {Setup}
\label{ansatz}
\label{section1}
 Let us show how the screening effects regulated by $\kv$ come about in a typical setup. Consider a Lagrangian made up by the standard GR and matter content plus an additional scalar (split into standard kinetic term + interactions) directly coupled to matter:
\bea
\mathcal{L}\sim \mathcal{L}_{E\-H}+ \mathcal{L}_m +(\partial_{\mu}{\phi})^2 +\mathcal{L}^{\rm int}_{\phi}+\frac{\beta}{M_{\rm Pl}}\,\phi\,T_{m} \;.
\label{hello}
\eea
 Such a scenario naturally emerges in dark energy models as well as e.g. in the \textit{decoupling} limit of modified gravity theories \cite{Ondo:2013wka,Fasiello:2013woa}.

 The existence of strong derivative $\phi$ self-interactions is the key to screening dynamics. As soon as the non-linearities in $\mathcal{L}^{\rm int}_{\phi}$ are important, they too will contribute a non-negligible kinetic term and affect the canonical normalization of $\delta\phi$. In other words, the kinetic term has the form $Z(\bar{\phi}) (\partial{\delta\phi})^2$, with $Z\rightarrow 1$ only in the linear regime. Upon normalizing one finds $\delta\phi\sim \delta\phi_c /Z_{\rm int}$, where again $Z_{\rm int}$ depends on the background value of $\phi$ and the self-interactions coefficients {(see e.g. \cite{deRham:2014zqa})}. For $Z_{\rm int}\gg1$, the field $\phi$ coupling to matter is heavily suppressed. We identify the condition $Z_{\rm int}\gg1$ with a strong screening regime where the presence of the additional dof will not be detectable \footnote{Although see \cite{Hui:2012jb}.}. 

 Let us schematically write the equation of motion for the dark matter + additional dof system in the Newtonian limit:
\begin{align}
\frac{\p \delta_m}{\p \tau}+\p_i [(1+\delta_m) v_m^i]=0\; , \;\,\,\,\, \quad\qquad\,\qquad\qquad\qquad\nonumber\\
\frac{\p v_m^i}{\p \tau}+ \mathcal{H}v_m^i+v_m^j \p_j v_m^i  =-\nabla^i  \Phi \; ,   \qquad\,\,\,\,\,\,\;\;\,     \qquad\quad \,\;\;  \nonumber\\ 
\nabla^2 \Phi= \frac{3}{2}\mathcal{H}^2\Omega_m \delta_m +F(\bar{\phi}) \nabla^2\delta\phi \qquad \qquad \qquad \qquad\, \nonumber \\
\nabla^2 \delta\phi +{\rm non\,linearities}= \frac{\beta}{M_{\rm Pl}}\,\delta_m  \;,\;\,
\qquad\qquad\qquad\;\;
\label{full}
\end{align}
where we have split the scalar background value from its fluctuations in  $\phi=\bar{\phi}+\delta\phi$ and used the fact that the fifth force from the extra field will affect dark matter dynamics via the Poisson equation. The function $F(\bar{\phi})$ tracks the screening strength and is therefore related to $Z_{\rm int}$.\\  Note that we have instead been deliberately agnostic about the equation of motion for $\delta\phi$: for stability, we require it be at most second order in time derivatives. A well-studied \cite{Bartolo:2013ws} example is the cubic Galileon:
\bea
\nabla^2 \phi+ \frac{1}{\Lambda^3} \Big[ (\nabla^2 \phi)^2 - (\nabla_i \nabla_j \phi)^2\Big]=\beta \frac{\rho}{M_{\rm Pl}} \; .
\label{g3}
\eea
 Galileon interactions are ubiquitous: one can think of them in this context as emerging in the decoupling limit of massive (bi)gravity or as a small subset of Horndeski-type interactions. Indeed, it has been shown \cite{Koyama:2013paa} that the broader class of Horndeski theories exhibits Vainshtein screening. The presence of $\Lambda$ in Eq.(\ref{g3}) identifies the threshold in momentum at which non-linearities become relevant. For example, in massive gravity $\Lambda=\Lambda_3\equiv (m^2 M_{\rm Pl})^{1/3}$, proving how a small mass can in principle activate screening at arbitrarily small momentum scales. However, if massive gravity is enlisted to explain cosmic accelaration, the value of $m$ cannot stray too far from the current value of the Hubble constant $H_0$. 

 Shifting the focus back on Eq.(\ref{full}), we identify the two regimes in the DE/MG side via $Z_{\rm int}$, with $Z_{\rm int}\ll1$ corresponding to the regime where the dynamics is well approximated by the linear solution, and  $Z_{\rm int}\gtrsim 1$ requiring non-linearities to be taken into account.

\section {Modelling Screening}
\label{filter}
 We want to model the observables resulting from the solution to Eq.~(\ref{full}) in a regime sensitive to screening effects. To this aim, we assume that the system has been solved up to a certain perturbative order ``$l$" in $k/k_{\rm NL}$ \footnote{The structure of the perturbative expansion is more complex in non-scaling universe but we nevertheless adopt it here for the sake of convenience.} and, in particular, that the solution is known for the \textit{total} density contrast variable defined as the RHS of the Poisson equation, $\nabla^2 \Phi \equiv\frac{3}{2}\mathcal{H}^2 \Omega_m\delta_T$. 

We model screening dynamics on crucial observables, such as the power spectrum of the total density, in the following way: 
\begin{align}
P_{\rm res} \big |_N (k,\tau) &= \sum_{n=0}^N P_{\rm res}^{(n)} (k,\tau)  \nonumber\\
&= \sum_{n=0}^N \int \frac{d^3 k'}{(2\pi)^3} ~\mathcal K^{N}_n(k',k,\tau) P^{(n)}(k^{\prime},\tau) ,
\label{resum}
\end{align}
 where $N$  stands for the perturbative order \textit{up to} which the expression is valid and $n$ signals instead a specific order in the expansion. We formally introduced here the kernels $\mathcal K^{N}_n(k^{\prime},k,\tau)$ to describe the resummed dynamics of higher order contributions in $k/k_{\rm V}$. In other words, kernels are to account for the part of the screening dynamics that is \textit{not} captured by the perturbative expansion. Indeed, the non-linearities in the DE/MG sector play an increasing role at higher momenta; given the hierarchy, the $k/k_{\rm V}$ parameter becomes order one much sooner than $k/k_{\rm NL}$ and so needs to be resummed. This resummation affects in particular also observables at $k\ll k_{\rm V}$. This further implies that kernels, in addition to varying according to the perturbative order index $n$, should also depend on the overall PT order $N$. The reason is that depending on the working PT order $N$, part of the screening is captured perturbatively, while the kernels are responsible for the resummation of the ``residual" screening. As we go higher in perturbation theory, the kernels have indeed less screening to account for.

The structure of Eq.~(\ref{resum}) is reminiscent  of the recently proposed resummation schemes in the context  of the baryon acoustic oscillations (BAO) \cite{Senatore:2014via,Vlah:2015sea,Blas:2016sfa} {(see also \cite{delaBella:2017qjy})}. The physics  we are describing is of course quite different but the analogy stems from the fact that here too the kernels account  for the effects from non-linear physics (in the  $k/k_{\rm V}$ expansion) that we need to resum. More specifically, in our case the non-linearities to be resummed as k approaches $k_{\rm V}$ are those in the dark energy/modified gravity sector. These are propagated to the dark matter sector gravitationally as clear from Poisson's equation. The coupling between the two sectors is suppressed as DE/MG non-linearities become important and in particular the contribution from the DE/MG sector to the total density contrast becomes much weaker. As a result, the kernels in Eq.~(\ref{resum}) are  sensitive to the DE/MG contributions to observables and essentially blind to the purely dark matter $\Lambda$CDM-like sector. 

A top-bottom exact derivation of the kernels, ideally via a Lagrangian formulation, is beyond the scope of this paper and we leave it to upcoming work. From here on instead, we proceed phenomenologically. Organizing the total power spectrum contribution in generalized cosmology, $P$, as a $\Lambda$CDM piece plus the remaining  $\Delta P =  P - P_{\Lambda\rm{CDM}}$, we can write, for the linear calculation,
\begin{align}
P_{\rm res}^{(0)} (k,\tau) & = P^{(0)}_{\Lambda\rm{CDM}} (k,\tau) + K_0 (k,\tau) \Delta P^{(0)} (k,\tau),
\label{47}
\end{align}
where $P^{(0)}_{\Lambda\rm{CDM}} $ is the usual linear $\Lambda$CDM power spectra (the usual output of Boltzmann codes such as CAMB \cite{Lewis:1999bs} or CLASS \cite{Lesgourgues:2011re,Blas:2011rf}) , while $P^{(0)}$ is the linear solution in generalized cosmology (also linear or low-order in the $k/k_{\rm V}$ expansion).  Note also that we shall refer to the total power spectrum also as $P_{\rm res}$, this to underscore the resummation of screening effects. The phenomenological nature of Eq.~(\ref{47}) is already evident from the fact that the kernels now act directly on the ``external" observables, as opposed to the more general prescription in Eq.~(\ref{resum}).
The kernel $K_0$ is in Eq.~(\ref{47}) to capture screening effects much beyond the linear order in the $k/k_{\rm V}$ expansion, it is a resummation to all orders in PT.
As one proceeds beyond linear order, the expression for the total power spectrum reads, after resumming the $k/k_{\rm V}$ expansion, as
\begin{align}
P_{\rm res} \big |_N (k,\tau) &= \sum_{n=0}^N \Big[  P_{\Lambda\rm{CDM}}^{(n)} (k,\tau) \nonumber\\
& \hspace{2.0cm} + K^N_{n}(k,\tau) \Delta P^{(n)} (k,\tau) \Big].
\label{eq:Pres2}
\end{align}
Here $P_{\Lambda\rm{CDM}}^{(n)}$ represents the $n-$th loop expansion of the power spectrum in the $\Lambda$CDM cosmology and $\Delta P^{(n)}$ perturbatively (both in the $k/k_{\rm V}$ and $k/k_{\rm NL}$) captures the dynamics beyond $\Lambda$CDM at every loop.
Let us stress again the effect of the $K^{N}_{n}$ factors, where the index $N$ stands for the PT order we are working at and the index $n$ stands for an expansion in $k/k_{\rm V}$: as one goes higher in perturbation theory (increasing N), more of the screening dynamics is captured already perturbatively and so the $N$ dependence of kernels $K^{N}_n$ is there to ensure one does not ``double count" the perturbative and the resummed screening contributions.  We will illustrate this below with a specific example. 

The discussion so far has been relatively general as the specific phenomenon we want to describe is encoded in the form of the kernels. We now specialize the analysis to the screening mechanism known as Vainshtein screening (VM). 

\subsection*{Vainshtein screening, -resummation-}
 In the mechanism first studied by Vainshtein \cite{Vainshtein:1972sx}, the suppression of the coupling with matter originates from kinetic interactions in the DE/MG sector, such as the ones generating the second term in Eq.~(\ref{g3}). We refer the reader to \cite{Lue:2004rj,Koyama:2007ih} for important early works in the context of structure formation {and to e.g. \cite{deRham:2011by,Taruya:2013quf,Taruya:2014faa} for more recent studies}. Let us see how the framework we have outlined takes shape in the case of VM.

\begin{figure*}[t!]
\label{fig2}
\includegraphics[scale=0.7]{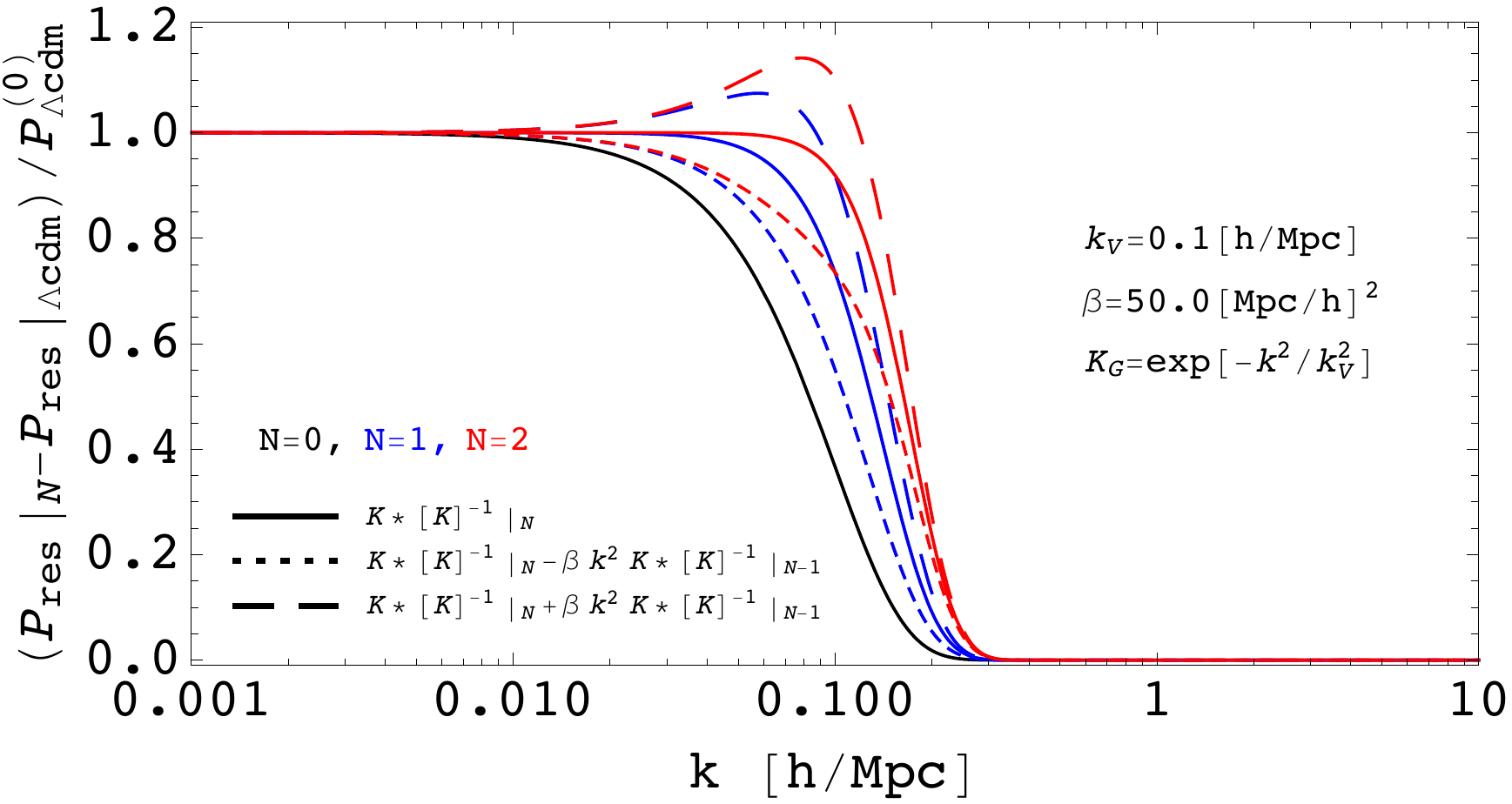}
\caption{The fractional difference between a fully screened total power spectrum and the $\Lambda$CDM result. Different colors indicate different perturbative orders. Continuous, dotted, and dashed lines stand for the action of the additional kernel $\tilde K$ in Eq~(\ref{23}), which in this approximation is regulated by the value of just one parameter, $\beta$. Note that for the $N=2$ case we have dropped the last contribution from Eq.~(\ref{23}) as it does not modify the physical picture in this specific example. }
\end{figure*}

It is convenient at this stage to express the generic $K^N_n$ in terms of the Taylor expansion of the generic reduced non-linear form $K^N$, i.e. we can write 
\begin{align}
K^N_n(k,\tau) = K (k,\tau) \big[ K \big]^{-1}\Big|_{N-n}  (k,\tau),
\label{ker1}
\end{align}
where the last term is the $(N-n)-$th order Taylor polynomial of the inverse of the reduced kernel $K^N$, typically a function of time and the $k/k_{\rm V}$ parameter. Note that in employing this form for the kernels there is already an element of choice. We now take on the form that the reduced kernels should have to account for Vainshtein screening. The most immediate constraints come from the asymptotic regimes:\\
- in the $k\rightarrow 0$, and for very low k in general ($k\ll k_{\rm V}<k_{\rm NL}$), these kernels (i.e. resummation) will not be necessary and must therefore reduce to unity.\\
- in the complementary regime, $k\gtrsim k_{\rm V}$ kernels ought to screen very efficiently and should therefore render any non-$\Lambda$CDM feature in the spectrum negligible.\\
The most natural candidates as reduced kernels $K(k,\tau)$ to model the VM are: 
 
\begin{align}
K_{\rm G}(k,\tau) &= \exp \left(- \sum_m \alpha_m (k/k_{\rm V})^{2m} \right), ~~ ~~ \nonumber\\
K_{\rm L}(k,\tau) &= 1/\left(1 + \sum_m \alpha_m (k/k_{\rm V})^{2m} \right),
\label{kernels}
\end{align}
where subscripts G and L indicate respectively Gaussian and Lorentzian forms. The presence of only even powers of $k$ is due to rotational invariance. Both expressions clearly satisfy the asymptotic requirements but the following considerations point to utilizing the Gaussian kernels.  From Eq.(\ref{ker1}) one can see how, in the case of the Lorentzian kernel, it is necessary for the sum over $m$ to go up to $\lfloor N/2 \rfloor+1$ in order to ensure that $K^N_n$ gives the desired asymptotic behaviour in the high $k$ limit. This in turn makes the reduced Lorentzian kernel $K_{\rm L}$ sensitive to the specific PT order one is working at. As a consequence, one should in principle write it with an $N$ index as well. Note that this is not the case for the Gaussian kernels. % for which the dependence on $N$ could come only through $\alpha_m$ terms.
Using the simplified form of Eq.~(\ref{ker1}), the formula in Eq. \eqref{eq:Pres2} becomes 
\begin{align}
\label{eq:Presaprox}
P_{\rm res} \big |_N (k,\tau) &= P_{\Lambda\rm{CDM}} \big |_N (k,\tau)  \\
& \hspace{0cm} +  K (k,\tau) \sum_{n=0}^N \big[ K \big]^{-1}\Big|_{N-n}  (k,\tau) \Delta P^{(n)} (k,\tau), \nonumber
\end{align}
 where again the form we have chosen for the kernels in Eq.~(\ref{kernels}) guarantees the correct behaviour in the asymptotic regions, with the second line of Eq.~(\ref{eq:Presaprox}) becoming negligible at sufficiently high $k$. 

\subsection*{Vainshtein screening, -perturbative build-up-}
 The framework that we have setup so far will account for the ``residual" screening effects, those that escape perturbation theory at the given working order. However, it is often the case the perturbative solutions themselves are hard to obtain without resorting to idealized configurations such as, for example, those endowed with spherical symmetry. On the other hand, an analytical handle on LSS dynamics is crucial in view of upcoming data from astronomical surveys. It is paramount that we develop analytical tools to complement the role of N-body simulations {(see \cite{Fang:2017daj} for interesting recent developments)} in the study of structure formation. 

In this context, the use and extension of Einstein-Boltzmann solvers to include DE/MG is an important and timely development \cite{Hu:2013twa,Zumalacarregui:2016pph} (see also \cite{Lagos:2016wyv}). However, it is often the case that available codes account only for the dynamics up to quadratic order in the Lagrangian and therefore do not fully account for screening. Our framework has already been setup to include the resummed screening component and we will now extend it to model also the perturbative screening build-up. This of course with the ultimate goal to make contact with simulations. 

As ever, the known behaviour in the asymptotics will act as our guiding principle. 
Let us proceed by assuming that we know the perturbative expression for the power spectrum up to order $n-1$ and would like to estimate the $n$-th order contribution. Since we organize our observables around the known $\Lambda$CDM result, the quantity to be determined at order $n$ will be  $\Delta P^{(n)}(k/k_{\rm NL},k/k_{\rm V}, \tau)$. In the following, we propose two different ways to estimate $\Delta P^{(n)}$. The first will be particularly effective at very low perturbative orders, as close as possible to the linear solution, the other in the complementary regime.
We first use the result (see e.g. \cite{Blas:2013aba}) valid for $\Lambda$CDM cosmology  expansion at large scales:
\begin{align}
P_{\Lambda{\rm CDM}}^{(n)} (k,\tau) / k^2 P_{\Lambda{\rm CDM}}^{(0)} \sim {\rm const}_{\Lambda}\,, ~~ k\to 0\,. 
\end{align}

%where we assume here that this expansion is still valid at scale beyond $k_{\rm V}$. 
It has been shown that this approximation is reliable up to scales of almost 0.1$\,{\rm h/Mpc}$ for the two-loop power spectrum and past 0.1 ${\rm h/Mpc}$ at higher orders (see Fig.~4 in \cite{Blas:2013aba}). We now extend this expansion beyond $\Lambda$CDM and write

\begin{align}
\label{51}
\Delta P^{(n)}(k,\tau)/ k^2\,  P^{(0)} \sim  {\rm const}-{\rm const}_{\Lambda } \frac{\, P^{(0)}_{\Lambda {\rm CDM}}}{ \, P^{(0)}} \;,
\end{align}

 where this is valid for $ k\ll k_{\rm V}$. The above equation provides our first estimate of the difference between the (unknown) perturbative expression at $n$-th order of the total power spectrum and the $\Lambda$CDM one. As such, this expression can be used in Eq.~(\ref{eq:Presaprox}) in order to include also the residual screening. 

In particular, whenever \footnote{Although this depends on the specific cosmology, one obvious parameter is the deviation of the equation of state from $w=-1$. More in general, one needs to establish to what extent going beyond-$\Lambda$CDM affects the variance of fluctuations and velocity dispersion. Once this  is done at one loop order, the results of \cite{Blas:2013aba} point to a reliable extrapolation to higher orders as well.} one can write $\Delta P^{(n)}(k,\tau)/ k^2\,  \Delta P^{(0)} \sim  {\rm const}$ and $k_{\rm V}$ happens to be small, e.g. $\sim 0.1\,{\rm h/Mpc}$, there is a dramatic simplification of the overall results for Eq.~(\ref{eq:Presaprox}), which in this case reads:
\begin{align}
\label{23}
P_{\rm res} \big |_N (k,\tau) &= 
P_{\Lambda\rm{CDM}} \big |_N (k,\tau) \\ 
&\hspace{-.8cm}+ K(k,\tau) \big[ K \big]^{-1}\Big|_{N}  (k,\tau) \Delta P^{(0)}(k,\tau)\nonumber\\
& \hspace{-.8cm}+K(k,\tau)\,\big[ K \big]^{-1}\Big|_{N-1}  (k,\tau)\, \Delta P^{(1)}(k,\tau) \nonumber \\
& \hspace{-.8cm} +k^2  \Delta P^{(0)} (k,\tau)  K (k,\tau) \sum_{n=2}^N \beta^0_n \big[ \tilde K \big]^{-1}\Big|_{N-n}  (k,\tau)\, , \nonumber
\end{align}
where we have used the fact that, for $k_V$ in the vicinity of 0.1$\,{\rm h/Mpc}$, one need only have the exact perturbative solution up to one loop and can rely on the approximation for higher orders contributions {(see e.g. \cite{Blas:2013aba} Fig 4.)}. {Note  that $\beta_n$ does in principle also depend on $k/k_{\rm NL}$.}  We stress that in this configuration the last term in the last line of Eq.~(\ref{23}) can be  further simplified in favour of the usual $K$ kernel times another compact kernel with no need for the sum over $n$. In Fig.~(2) we illustrate how the fractional difference between a fully screened total power spectrum and the $\Lambda$CDM PS would look like whenever the relation in Eq.(\ref{51}) can be simplified this one step further. In particular we assume the following $\Delta P^{(0)} \sim\,P^{(0)}_{\rm \Lambda CDM}$ and $\Delta P^{(n)} \sim  k^2 \Delta P^{(0)} \sim k^2  P^{(0)}_{\rm \Lambda CDM}$ .

Let us now consider another way to estimate the $\Delta P^{(n)}$ and its embedding in Eq.(\ref{eq:Presaprox}):
\begin{align}
\label{21}
P_{\rm res} \big |_N (k,\tau) &= P_{\Lambda\rm{CDM}} \big |_{N} (k,\tau) \\
& \hspace{0cm} +  K (k,\tau) \sum_{n=0}^{N-1} \big[ K \big]^{-1}\Big|_{N-n}  (k,\tau) \Delta P^{(n)} (k,\tau)\nonumber \\
& \hspace{0cm} +  K (k,\tau)\, \bar{K} (k,k/k_{\rm NL},\tau)\, \bar{\Delta}P^{(N)} (k,\tau)\,, \nonumber
\end{align}
where in the last line we have isolated the term $\Delta P^{(N)}$ to be estimated and written it as $\Delta P=\bar{K}\bar{\Delta}P$. The role of the new kernel $\bar{K}$ is to model the perturbative screening contribution and that is why it must depend also on $k/k_{\rm NL}$. More explicitly, in order to estimate the value of $\Delta P$ at higher perturbative orders, we propose the following:
\begin{align}
\Delta P^{(N)}=  \bar{K} \bar{\Delta} P^{(N)} \equiv \bar{K} \Delta P^{(N-1)} ,
\label{22} 
\end{align}
 where we are using the fact that, at higher orders, the most reliable way to estimate $\Delta P^{(N)}$ is to employ the value of the known closest observable, $\Delta P^{(N-1)}$, and control it with $\bar K$. Let us then explore some of the properties we demand of the new kernel. First of all, at $k/k_{\rm NL}$ scales where the $N$-th order contribution in perturbation theory becomes relevant, we require that $\bar K\ll 1$ so that $\bar K\,\Delta P^{N-1}$ is effectively of order $N$.\\
\indent The specific form of $\bar K$ is hard to pin down for a generic theory with a screening mechanism that could be either perturbatively very strong or very weak at order $N$ in the expansion. However, the task becomes easier if the scales where the $N$-th order contribution is important are also the ones at which screening becomes \textit{rapidly} strong. In such a scenario, even if $\bar K$ is modeling a perturbative contribution to screening, the rapid perturbative onset of screening will be well-approximated by the  Gaussian or Lorentzian form in Eq.~(\ref{kernels}) and suitable $\alpha_n$ coefficients will readily account for an effect of order $N$ (and not $N-1$) in a $\Delta P$ derived via Eq.~(\ref{22}). Note also that in the rapid perturbative screening limit $\bar K$ need not depend on $k/k_{\rm NL}$: the dynamics of the two expansions decouples in this limit and at the next perturbative order one may well use directly the $\Lambda$CDM result for the total power spectrum.
 %eccolo3
 
%\noi {\rosso --------------------------------------------------------------------------}\\

\section{Embedding in the ``EFT of LSS"}
\label{embedding}
 The modeling of screening we have proposed can be readily embedded within the effective approach to LSS dynamics. Let us consider the asymptotic regions. For very small $k$ the shielding effect  is negligible and the EFT prescription \cite{Carrasco:2012cv} will generate the appropriate \textit{counterterms} for both the dark matter and dark energy component. We stress that at small perturbative order the counterterms can be common to both components \cite{Fasiello:2016yvr} or, in other words, degenerate (see \cite{Lewandowski:2016yce} for a derivation).
At perturbative orders above the one where (strong) screening occurs observables coincide with their $\Lambda$CDM counterpart and so do counterterm operators. This is natural as our filter is nothing other than a phenomenological resummation  in the $k/k_{\rm V}$ expansion; as such, it bypasses the need for counterterms. The coefficients $\alpha_n$ in our kernels will also vary depending on the different shielding strengths associated with different screening theories/interactions. Such difference can be found already within the same model: for example, the cubic, quartic and quintic Galileon interactions generate a different suppression of the coupling to matter.

\section{Conclusions}
\label{conclusions}
The data from upcoming astronomical surveys (Euclid, LSST) will put to the test our best ideas on the mechanism responsible for the current acceleration of the Universe. There are intriguing proposals that go beyond the $\Lambda$CDM model: from dark energy to IR modifications of general relativity. The additional degrees of freedom that typically characterize beyond-$\Lambda$CDM models come with an associated scale, $k_{\rm V}$, beyond which the corresponding fifth force is suppressed to the point of being currently undetectable.
As we have seen, if the hierarchy between $k_{\rm V}$ and the scale of dark matter non-linearities $k_{\rm NL}$ is benevolent, $k_{\rm V}\lesssim k_{\rm NL}$, screening dynamics will be accessible in LSS setups already at quasi-linear scales. This is precisely the regime where perturbative analytical tools, such as effective field theory, are most efficient.

In this work we have proposed a phenomenological ansatz to model screening dynamics. Our framework accounts for the ``residual screening" that is not captured in perturbation theory but is crucial to obtain reliable predictions for LSS observables. We have further put forward  a mechanism to estimate also the perturbative screening component whenever the exact result is not known. Our formalism can be readily adapted to several screening mechanisms and to different layers of approximation. In the second part of the text however, we have adopted to focus on one specific screening mechanism,  Vainshtein screening, and provided the corresponding kernels $K$.

We applied our formalism to the total density power spectrum, for which we have provided a  resummation scheme for higher order effects in $k/k_{\rm V}$. We stress in particular the usefulness of Eq.(\ref{23}): under certain assumptions it can model screening by relying on exact inputs solely from the linear theory.

Our approach here is phenomenological in nature. However, by enforcing a number of constraints from exact asymptotic solutions and from symmetries of the physical system, we have been able to identify very efficient kernels that account for screening dynamics. The most natural next step is to analyze screening via the Lagrangian PT formalism, which we address in upcoming work.

\section{Acknowledgements}
\label{acknowledgements}
\noi We are delighted to thank Diego Blas for illuminating conversations. We are very grateful to Martin White and Miguel Zumalacarregui for precious feedback on a draft version of the manuscript. MF is supported in part by NSF PHY-1068380; ZV is supported in part by DOE contract DEAC02-76SF00515.\\

\end{document}